# AN ELECTRONIC DIGITAL COMBINATION LOCK: A PRECISE AND RELIABLE SECURITY SYSTEM

Adamu Murtala Zungeru

School of Electrical and Electronic Engineering, University of Nottingham, Jalan Broga, 43500 Semenyih, Selangor Darul Ehsan, Malaysia
adamuzungeru@ieee.org

## ABSTRACT

*The increasing rate of crime, attacks by thieves, intruders and vandals, despite all forms of security gadgets and locks still need the attention of researchers to find a permanent solution to the well being of lives and properties of individuals. To this end, we design a cheap and effective security system for buildings, cars, safes, doors and gates, so as to prevent unauthorized person from having access to ones properties through the use of codes, we therefore experiment the application of electronic devices as locks. However, a modular approach was employed in the design in which the combination lock was divided into units and each unit designed separately before being coupled to form a whole functional system. During the design, we conducted Twenty tests with the first eight combinations being four in number, the next seven tests being five and the last five combinations being six. This was done because of the incorporation of 2 dummy switches in the combinations. From the result obtained, combinations 8, 11, 13 gave the correct output combination. However, 8 being the actual combination gave the required output. The general operation of the system and performance is dependent on the key combinations. The overall system was constructed and tested and it works perfectly.*

## 1. INTRODUCTION

Due to the advancement of science and technology throughout the world, there is a consequent increase in the rate and sophistication of crime [1]. As a result, it is necessary to ensure security of oneself and one's valuable belongings. Even with the use of mechanical locks, the crime rate still has increased due to the fact that these locks are easily broken. Consequently, there is a need for other types of locks especially electronic ones [2-4].

This work is on the design and construction of an electronic combination lock with a keyboard to be mounted on the door for keying in the secret code. The code unit, which operates with a 10-switch (non- matrix) keyboard was designed to control an electromagnetic door lock with a four– digit code. Unlike other keyboard combination locks this lock is constructed in such a way that once any of the wrong keys is pressed, it resets automatically making it harder for an intruder to break into [2,5].

The increasing rate of crime, attacks by thieves, intruders, vandals etc., despite all forms of security gadgets and lock constitute the main factor that prompts the selection of this design. Therefore, the main aims of the design are: (1) To design a cheap and effective security system for buildings, cars, safes, doors and gates etc., (2) To experiment the application of electronic devices as locks, and (3) To prevent unauthorized person from having access to ones properties through the use of codes. This research work is limited to the historical development of very large integrated (VLI) Circuit and the working principle of various models. Materials used for the construction of the Circuit were sourced and put together locally. Due to financial and time constraints a 4043 IC was used instead of programmable IC. As limitations for the design, the 4043IC is a data latch used for data storage, the data stored cannot be changed in the case of the code being compromised but the data can be changed in the programmable IC if there is any compromise.

For clarity and neatness of presentation, we outlined the article into five major sections. The First Section gives a general introduction of security systems and combinational lock. Section two gives presentation of related work. In Section Three, detail descriptions of the design and implementation procedures are presented. Section Four presents the experimental results and discussion of the results. In Section Five, we conclude the work with some recommendations. Finally, references used in the manuscript are presented at the end of the paper.

## 2. RELATED WORK

This section brings to light the historical development of the lock, the types and functions of locks. A brief review of electronic combination locks and a discussion of major components that are used in electronic locks are presented.

### 2.1. Historical Development

The earliest lock in existence is the Egyptian lock, made of wood, found with its key in the palace ruins in Nineveh, in ancient Assyria [4-5]. In the $19^{th}$ century, level locks, cylinder locks and keyless locks were invented and improved upon [3]. The first successful metal key changeable combination lock was invented by James Sargent in 1857 [6]. This lock was the prototype of those being used in contemporary bank vaults.

In 1958, the first electronic combination lock was invented [3,7]. As subsequent developments were along the lines, the locks were improved upon by the improvement of materials and increasing complexity of the working mechanisms including the increasing use of automatic electronic alarm and safety devices [2,8].

### 2.2. Types and Functions of Locks

There are many types of lock that are in existence in our world today of which the main types are: (1) Mechanical Key locks, (2) Magnetic locks, and (3) Electronic locks.

1. Mechanical Key Locks: these are locks that consist of a bolt that may be slid to and fro, or rotated by a key (e.g. Padlock). In these types of locks, there are obstacles called wards or tumblers that permit only the right key to be turned on. This is mostly applicable in doors, gates and windows of houses, stores etc. A key and its hole in the shock shown in Fig. 1.

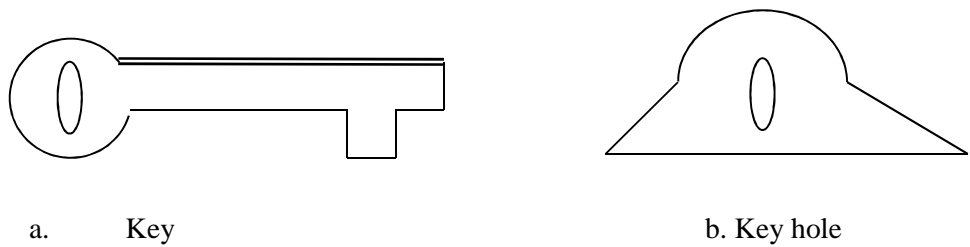

a.    Key                                         b. Key hole

**Fig. 1:** Diagram of a Mechanical Key Lock

2. Magnetic Lock: These are locks that are operated based on the theory of magnetism. These types of locks consist of bolts connected with magnets to ensure that they are locked. The key (which is usually a ferrous metal foil) when inserted pulls the bolts thereby releasing the lock to ensure it is opened (e.g. Solenoid). It is mainly used in residential as well as administrative areas (e.g. Offices). This mechanism is illustrated in Fig. 2.

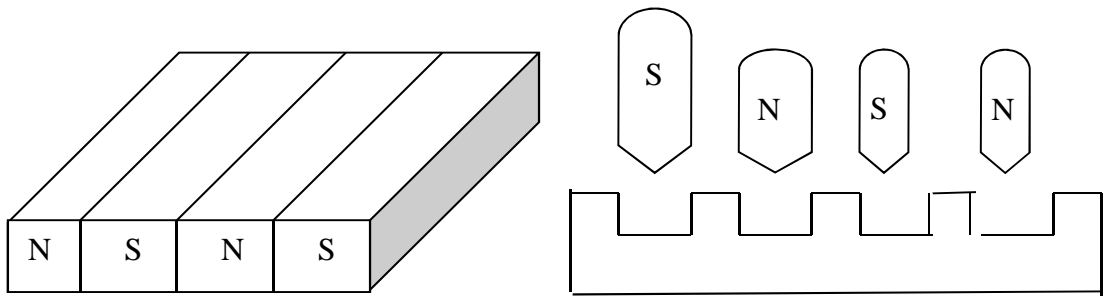

**Fig. 2:** Diagram of a Magnetic Lock

3. Electronic locks: these are the most sophisticated of all the locks [6]. This is because unlike the two other types that require a sort of key to open them, this doesn't require a key so that the issue of keys getting missing does not come into play. These types of locks are driven basically by electronic means and they are used mostly in industrial areas and areas where a high level of security is needed. Some of the electronic locks are:

   i. DNA Sensor Locks: these are electronic locks that compute into their memories, the genetic make-up of the individual such that only individuals that have their DNA computed into its memory would be allowed to enter. This is used in foreign countries in places like the Pentagon where a high level of security is needed.

   ii. Card sensor Locks: These are electronics locks that use the cards as keys such that when the card if inserted, it generates voltage by closing the circuit and energizing the relay which then opens the lock. These are used in industrial areas. This is shown in Fig. 3.

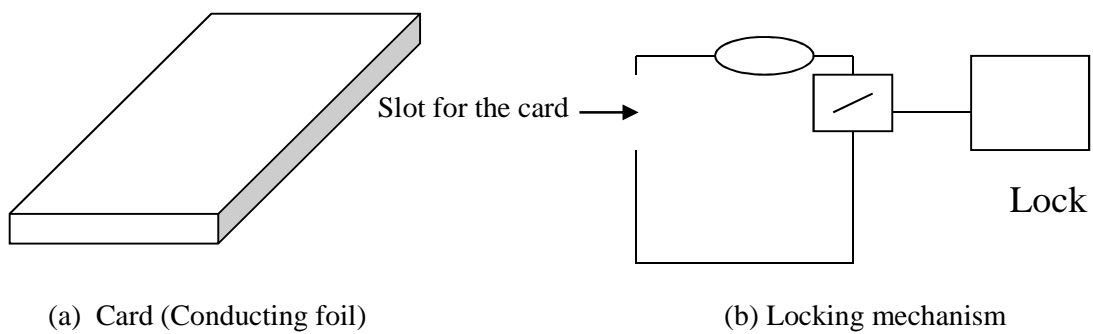

 (a) Card (Conducting foil)       (b) Locking mechanism

**Fig. 3:** Diagram of an Electronic Card Sensor Lock

   iii. Electronic Eye Locks: these are electronic locks that compute into their memories, the picture of the individual's eye such that only people with their eye pictures in the memory in the memory who will be allowed to enter.

   iv. Thumbprint sensor locks: these are electronic locks that use the thumbprint of the user as the key such that it is only individuals whose thumbprints have been inputted in its memory would the lock open for.

   v. Electronic Combination Locks: these are electronic locks that are operated by inputting the correct code by means of an external device such that only people that know the code can open the lock.

Though in this work, much attention is given to electronic combination lock as it is the subject of the design and implementation of an Electronic Digital Combination Lock: A Precise and Reliable Security System.

## 2.3. Electronic Combination Locks

The use of electronic combination locks in modern day technology cannot be overemphasized. They prevent losses to theft, carelessness e.t.c. Also being that they are electronically powered, they provide a better and safer security system than all other locks [2].

However, the electronic combination lock though very good has some drawbacks which are: (1) being that they are powered by electronic means, they are susceptible to power failure. To avoid this, the circuit needs to use both AC main supply and also D.C. supply from a battery or any other source. (2) Since it has to be mounted at a point outside the building, a lot of undesirables may be found tapping on the buttons and even resulting in the damage of the control unit mounted with it. To avoid this, the control units need to be placed away from the input unit.

## 3. DESIGN THEORY OF THE SYSTEM

This section deals with the design procedure for the electronic combination lock. The block diagram of the power supply unit, the input, control and the output units are shown in the figure below (Fig. 4).

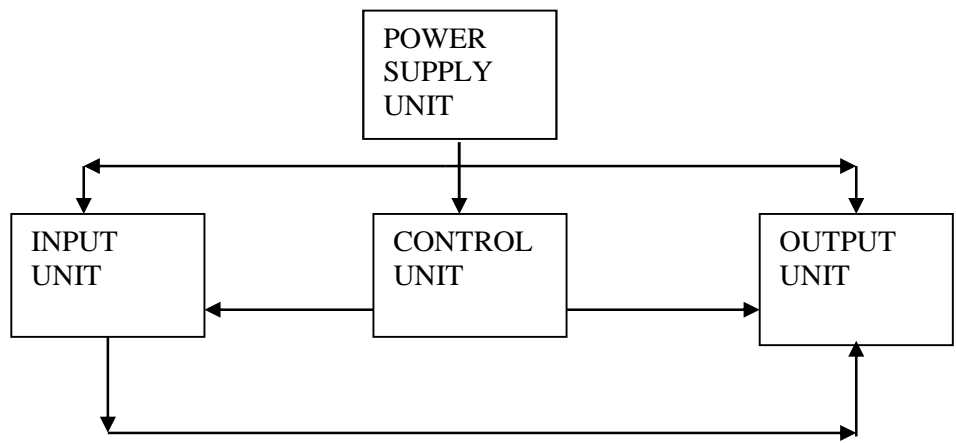

**Fig. 4**: Block Diagram of the Electronic Combination lock

## 3.1. The Input Unit

The input unit comprises mainly of the keyboard and its switches each can generate a discrete signal when processed. It is made up of ten switches, of which four will be used as the key in the secret code, another four will serve as the reset switches, and the remaining two will serve as a decoy.

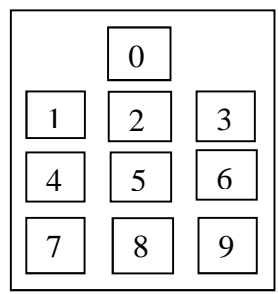

**Fig. 5:** The Keyboard

By using the ten switches, the number of possible combinations was calculated. By applying the permutation principle, the numbers of entries that can be made is given by;

$$P = {}^{10}C_4 \quad (1)$$

$$= \frac{10!}{6!4!}$$

$$= \frac{10 \times 9 \times 8 \times 7 \times 6 \times 5 \times 4 \times 3 \times 2}{6 \times 5 \times 4 \times 3 \times 2 \times 4 \times 3 \times 2}$$

$$= 210$$

This means that there are 210 ways in which this combination can be set, which means that the probability of an intruder to break the code is 1 out of 210 ways. The keys that set up the code are 9,5,0,2, which are switches S1 to S4 in Fig. 5 and 6 respectively. Each of these switches is linked to a bi-stable (flip-flop). The reset switches are 3,4,7,8, while keys 1 and 6 are the decoy. The figure below show the details of the input unit.

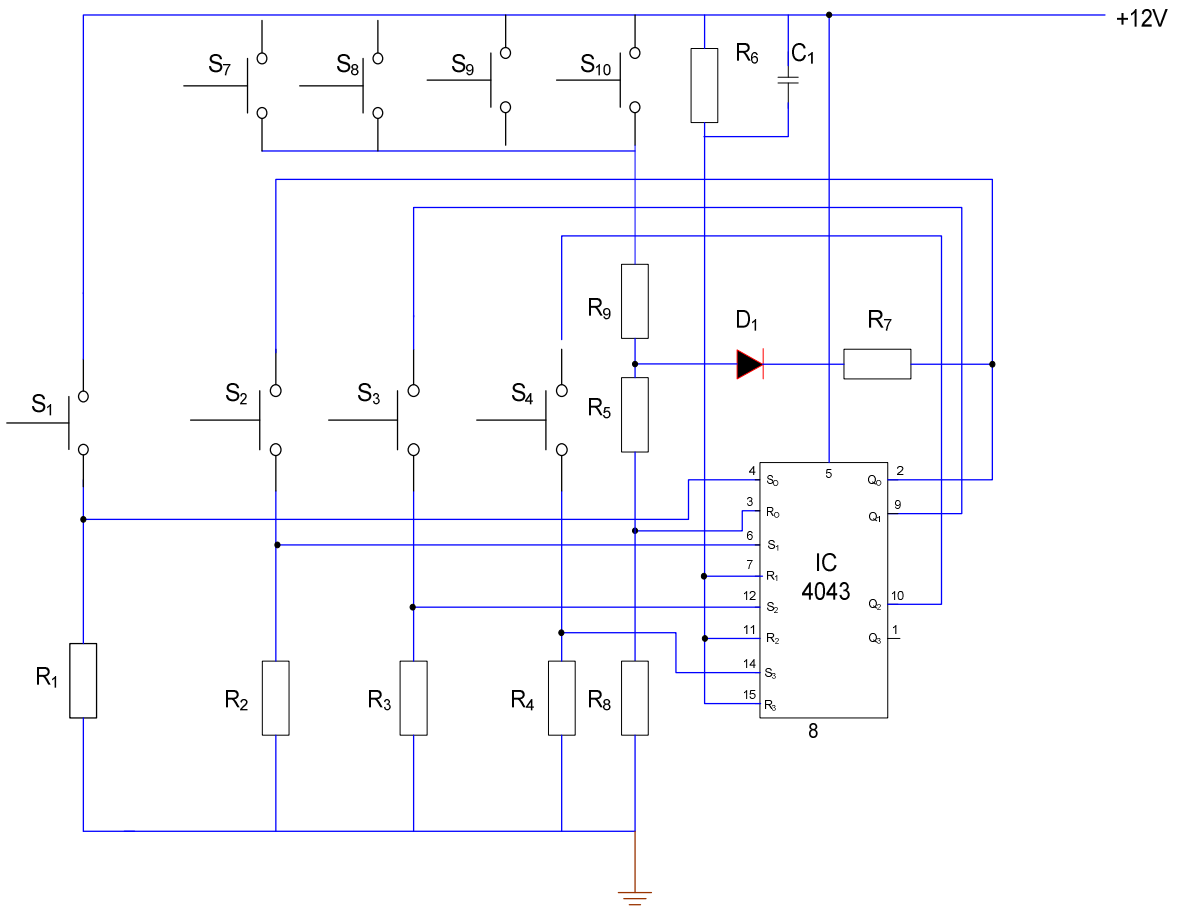

**Fig. 6:** The input and control unit

Whenever the keys are pressed in a sequential order, a Q-output of logic 1 or HIGH is the expected output. Voltage supply is made available to the switch (S1). They are grounded via pull-up resistors (i.e. R1, R2, R3, & R4) in order to fix the voltage approximately to the voltage supply of 12V. For this to be done the resistance of the pull-up resistors must be high enough to ensure that there is a voltage drop across the resistor to ground. Hence the value of each R1, R2, and R3 and R4 are chosen to be 10k Ω.

The combination switches are connected to the set input of their respective bistables such that for there to be a logic HIGH at Q-output , the set input of the bistable must be high. The dummy switches are not connected in the circuit so they act as a decoy to any burglar who tries to break in the lock.

The RC circuit forms a debunking network whose function is to perform a timing operation of the circuit. In designing it, a time of 4.7seconds was chosen which is long enough for the user to open the lock and then lock it again. The formula is given by:

$\tau = R_6 C_1$ (2)

$C_1$ was chosen to be 10μF, $\tau = 4.7s$

Therefore, $R_6 = \frac{\tau}{C} = \frac{4.7}{10 \times 10^{-6}} = 470000 \Omega$

$R_6 = 470 K\Omega$.

The resistors $R_9$ and $R_5$ form a voltage divider network for the output $Q_0$ when at rest. The resistors $R_5$ and $R_8$ form a voltage divider network for the rest input R0.

The diode D1 was chosen to be IN4148 because of its switching ability which is necessary in the circuit to prevent current reversal when the current is reset either via the reset switches or the debouncing network.

Before key S1 is operated, the Q output of the first bistable is low so that the reset input is held low via R1, D1, and R5. The potential across C1 is then almost equal to the voltage supply (i.e. 12V). Because of D1, the capacitor will discharge slowly due to the fact that D1 offers a resistance to the current flow that is generated by the voltage of capacitor C1 and network by R5 and R8.

## 3.2. The Control Unit

When a key is pressed, the control unit receives signals from the input unit and processes them. The control unit comprises of a QUAD three state two cross-coupled NOR gates. The RS latch when SET and RESET are both high, the output is unpredictable. This implies that for the combination lock to work as expected, this condition should be avoided and this is done by connecting the set inputs to the push-button switches (i.e. S1 toS4) while the reset inputs are connected in series with the debouncing network that is responsible for the timing the circuit when it is still ON.

The operation of the control unit is such that when the first key is pressed, the set input S0 is HIGH so that the output Q0 is HIGH, as the other switches are pressed in the correct sequence, their respective outputs go HIGH and the relay is energized. Then, the set input goes low thereby leaving the output still HIGH for the period determined by the debouncing network (i.e. RC circuit). After that, the capacitor discharges its voltage into the reset inputs thereby making the reset inputs HIGH. This, in turn, brings the output to zero after which the reset inputs then change to zero and that confirms that the relay is de-energized.

## 3.3. The Output Unit

The output unit comprises of a relay which drives the solenoid and a transistor (connected to a switch) which is used to interface the RS output Q3 from the control unit and the relay. The transistor T1 which was chosen was the BC457 and it was chosen to prevent the unregulated voltage from driving back into the IC output terminal thereby causing damage and also to increase the current because of the current demand of the relay. Also, the switching diode D2 (IN4148) was placed in parallel with the relay during the voltage fluctuation.

The transistor used is the BC547 with the following data; $V_{CC} = 12V$, $V_{BE} = 0.7V$, $R_B = 4.7KΩ$, $h_{FE} = 320$.

Using the equation below, we can find the collector current as follows:

$$V_{CC} - V_{BE} - V_{Q3} = 0 \qquad (3)$$

Using the values above, we have

$$12 - 0.7 - V_{Q3} = 0$$

$$V_{Q3} = 11.3V$$

But $V_{Q3} = I_B \times R_B$ \qquad (4)

That is, $11.3 = 4.7 \times 10^3 \times I_B$

Therefore $I_B = \frac{11.3mA}{4.7} = 2.4mA$

Also, $I_C = h_{FE}I_B$ \qquad (5)

$= 320 \times 2.4mA$

$= 0.769A$

But $I_E = I_B + I_C$ \qquad (6)

$= (0.769 + 0.002)A$

$= 0.771A$.

It can be seen from equation that the current flowing into the base of the transistor is small (i.e. 2.4mA) because of the value of the resistor (4.7KΩ). This is to ensure that the transistor does not burn as a result of high voltage across the output $Q_1$.

In the circuit, the double pole relay was used because it could switch the solenoid and lock and open indicators.

### 3.4. The Power Supply Unit

The power supply unit was bearing in mind that the CMOS logic IC requires a DC regulated voltage of 12V. The unit is made up of a step-down transformer, a bridge rectifier and a voltage regulator. Fig. 7 shows the circuit diagram of the power supply.

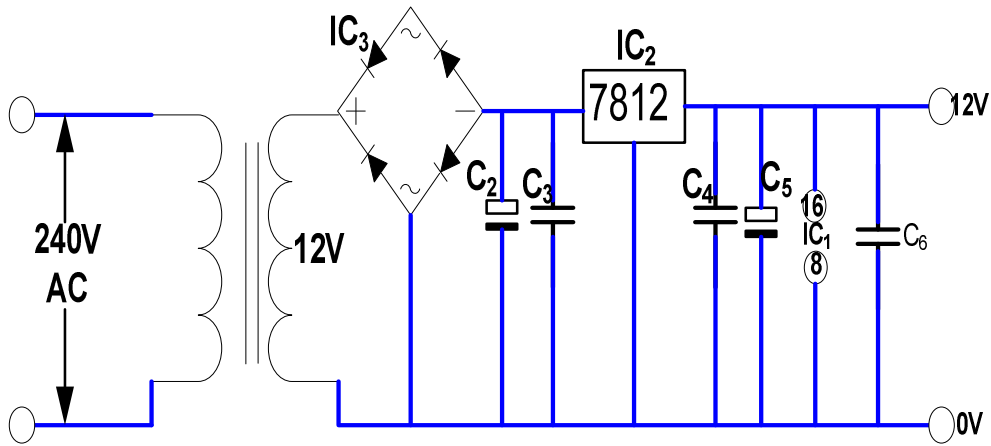

**Fig. 7:** The Power Supply Unit

The transformer is a step down type rated 500mA, 240V/12V. The a.c. Mains supply is applied to the primary winding, which is stepped down to 12V across the secondary winding.

A full wave rectification of this 12V is achieved using a bridge rectifier to allow for the current needed by the electronic circuit and to provide some allowance to improve the reliability of the circuit. It also occupies lesser space than four diodes would. The capacitor $C_2$ (shown in Fig. 7), which has a value of 470µF smoothens the ripple voltage from the rectifying diode output. $C_2$ is electrolyte because it is a power filtering capacitor. Normally, voltage regulator circuits are designed to give the specific output required to make the circuit function. For this design, the IC regulator 7812 was used which gives an output of +12V d.c. required by the electronic circuit. Fig 8 shows the circuit diagram of the regulator circuit.

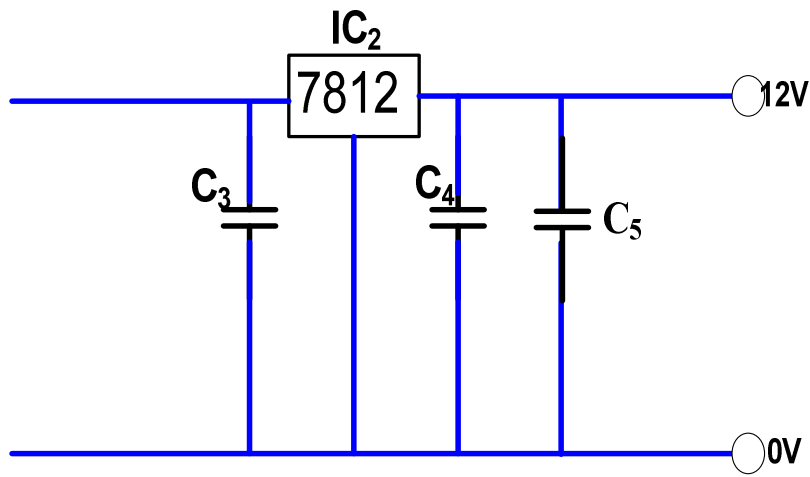

**Fig. 8:** The Voltage Regulator Circuit

The capacitors $C_3$ and $C_4$ which have capacitance of 100nF each are used to maintain the d.c. voltage and additionally assist in removing any high frequency component. For all 78 series, the 100nF capacitor is a standard component. The reservoir $C_5$ which is electrolyte in nature is used to assist the input and output capacitors of the voltage regulator to stabilize the regulated output.

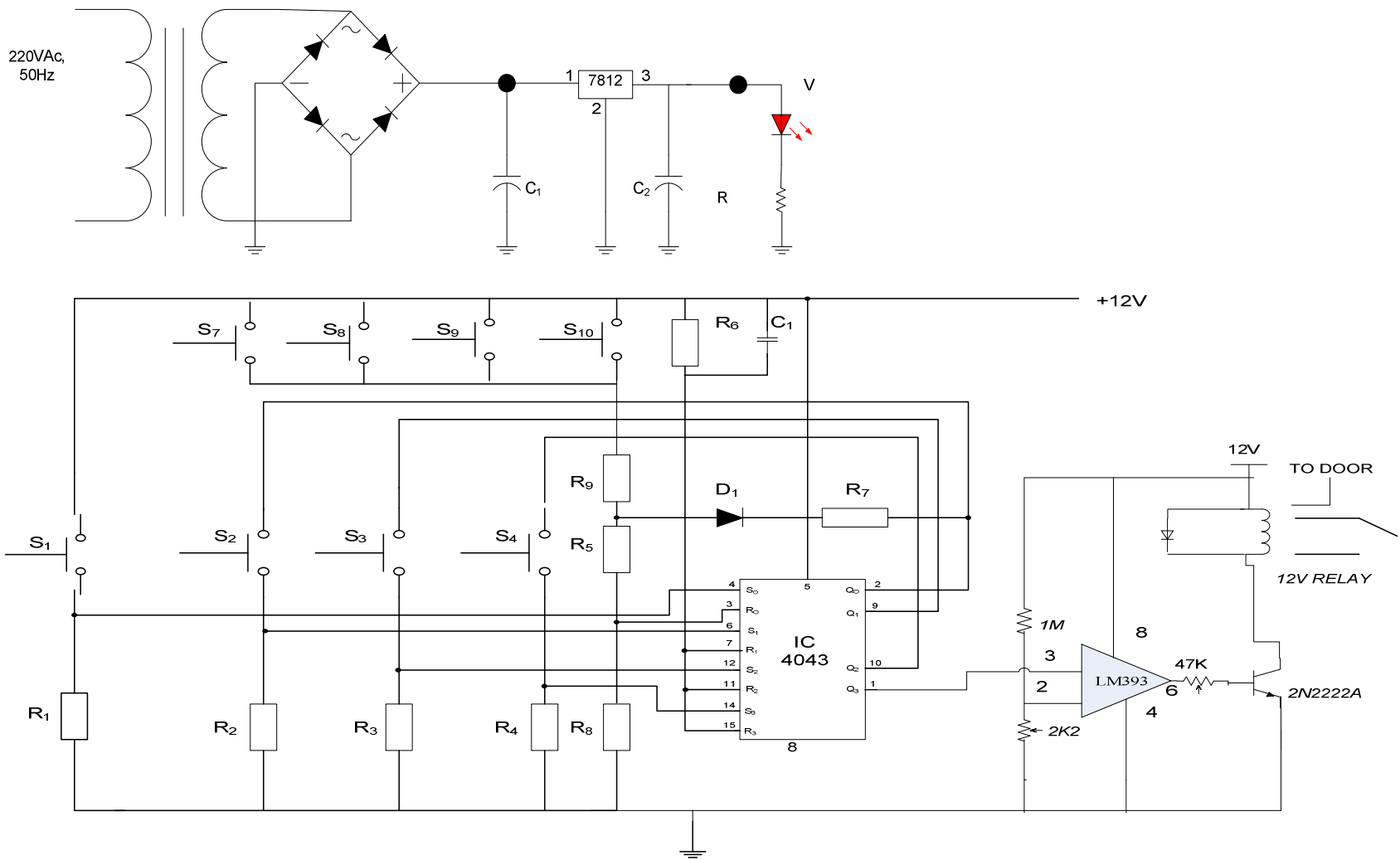

**Fig. 9:** Complete Circuit Diagram of the Electronic Combination Lock System

# 4. CONSTRUCTION, TESTING AND RESULTS

This section deals with the circuit construction, packaging, the test and the result obtained. The justification of the results obtained are also presented.

## 4.1. Construction

The control unit was first (temporarily) constructed on a breadboard with all components put in place to test how well they would function as a unit. This also made it easier in the layout as some wrongly positioned components were re-arranged for neatness. It was also tested using a +12V D.C power supply unit and found to be working as expected

In the layout of the components, certain factors were put into consideration. They include spacing, shape and size of the components, wires (jumpers) used and the number of components. The control circuit, the power supply circuit and the output circuit (less the solenoid) were constructed on the same Veroboard. The IC for MC14043 was mounted and soldered in position. The resistors, transistors and capacitors came next. The wires (jumpers) followed suit and then the relay. The contacts of the relay were then connected externally to the solenoid and another transformer T2 with 15V output. Special attention was paid to the pin connection of the transistor used. The input unit was connected externally from the control, output (less the solenoid) and power supply units with the wires connecting the switches properly soldered in position. Also, attention was paid to the polarity of the bridge rectifier and the capacitors $C_2$ and $C_5$. The power supply was tested on no- load using a 240V AC supply and voltage readings were taken at different points.

## 4.2. Packaging

The circuit was packaged in three Perspex cases with one of the Perspex cases exposed and carrying the solenoid. Between the input unit and the control unit are special wires connecting them called wire jumper. It was used because of its high flexibility and resistance to strain. The casing for the input unit has dimensions 98mm x 78mm x 30mm while that of the main circuit, which houses the control unit, power supply unit and the output unit (less the solenoid) has dimensions 230mm x 120mm x 80mm. The circuit board was laid out on the floor of the box with the transformer mounted beside it, while that of the input had the circuit board carrying the switches glued to the top of the casing with a hole large enough to contain the keyboard.

## 4.4. Tests Carried Out and Results Obtained

On completion of the construction, voltage readings (using a digital Multimeter) were taken at various points with both the right combination and the wrong combinations of numbers made. Various number combinations were made and the circuit was observed at various points. The deductions made based on this are tabulated and shown in Table 1. The HIGH output is 11.3V (i.e. $V_{Q1} = V_{Q2} = V_{Q3} = 11.3V$) while LOW output is 0.7V.

## 4.5. Discussions

Twenty tests were carried out with the first eight combinations being four in number, the next seven tests being five and the last five combinations being six. This was done because of the incorporation of 2 dummy switches in the combinations. From the result obtained, combinations (8), (11), (13) gave the correct output combination (8) being the actual combination gave the required output. The remaining combinations being that they were pressed with the dummy switches still gave a HIGH output because they do not affect the outcome. The remaining combinations however that gave LOW output were so either because of the wrong sequence of the combination or because the reset switches were pressed along with the combination thereby giving a LOW output.

**Table 1**: Various Number Combination and their Respective Outputs

| S/No | Different number combinations | Output | | | | | | |
|---|---|---|---|---|---|---|---|---|
| | | $Q_0$ | $Q_1$ | $Q_2$ | $Q_3$ | Solenoid | Green Indicator | Red Indicator |
| 1 | 1,7,5,2 | HIGH | LOW | LOW | LOW | CLOSE | OFF | ON |
| 2 | 1,5,9,7 | HIGH | LOW | LOW | LOW | CLOSE | OFF | ON |
| 3 | 1,9,2,2 | HIGH | HIGH | LOW | LOW | CLOSE | OFF | ON |
| 4 | 1,5,7,9 | HIGH | LOW | LOW | LOW | CLOSE | OFF | ON |
| 5 | 1,9,5,3 | LOW | LOW | LOW | LOW | CLOSE | OFF | ON |
| 6 | 8,4,9,3 | LOW | LOW | LOW | LOW | CLOSE | OFF | ON |
| 7 | 1,9,5,2 | LOW | LOW | HIGH | LOW | CLOSE | OFF | ON |
| 8 | 9,5,0,2 | HIGH | HIGH | HIGH | HIGH | OPEN | ON | OFF |
| 9 | 3,8,7,9,0 | HIGH | HIGH | LOW | LOW | CLOSE | OFF | ON |
| 10 | 1,9,5,3,5 | LOW | LOW | HIGH | LOW | CLOSE | OFF | ON |
| 11 | 9,5,0,1,2 | HIGH | HIGH | HIGH | HIGH | OPEN | ON | OFF |
| 12 | 2,1,9,5,6 | HIGH | HIGH | HIGH | LOW | CLOSE | OFF | ON |
| 13 | 9,5,0,6,2 | HIGH | HIGH | HIGH | HIGH | OPEN | ON | OFF |
| 14 | 1,9,0,5,3 | HIGH | HIGH | LOW | LOW | CLOSE | OFF | ON |
| 15 | 1,9,5,3,4 | HIGH | HIGH | HIGH | LOW | CLOSE | OFF | ON |
| 16 | 8,6,1,0,9,3 | LOW | LOW | LOW | LOW | CLOSE | OFF | ON |
| 17 | 1,2,9,6,5,3 | HIGH | HIGH | HIGH | LOW | CLOSE | OFF | ON |
| 18 | 1,4,9,2,7,3 | HIGH | LOW | LOW | LOW | CLOSE | OFF | ON |
| 19 | 3,7,5,1,3,6 | LOW | LOW | LOW | LOW | CLOSE | OFF | ON |
| 20 | 0,9,1,5,2,3 | LOW | LOW | LOW | LOW | CLOSE | OFF | ON |

## 5. CONCLUSIONS

The increasing rate of crime, attacks by thieves, intruders and vandals, despite all forms of security gadgets and locks still need the attention of researchers to find a permanent solution to the well being of lives and properties of individuals. As such, we design a cheap and effective security system for buildings, cars, safes, doors and gates, so as to prevent unauthorized person from having access to ones properties through the use of codes, we therefore experiment the application of electronic devices as locks. The system works by combination lock which was divided into units and each unit designed separately before being coupled to form a whole functional system. Twenty tests were conducted to ascertain the reliability of the design with the first eight combinations being four in number, the next seven tests being five and the last five combinations being six. This was done because of the incorporation of 2 dummy switches in the combinations. From the result obtained, combinations 8, 11, 13 gave the correct output combination. However, 8 as the actual combination gave the required output. The general operation of the system and performance is dependent on the key combinations. The overall system was constructed and tested and it works perfectly.

## ACKNOWLEDGEMENTS

The author would like to thank Col. Muhammed Sani Bello (RTD), OON, Vice Chairman of MTN Nigeria Communications Limited for supporting the research.

**Authors**

**Engr. (Dr) Adamu Murtala Zungeru** received his BEng in electrical and computer engineering from the Federal University of Technology (FUT) Minna, Nigeria in 2004, MSc in electronic and telecommunication engineering from the Ahmadu Bello University (ABU) Zaria, Nigeria in 2009, and his PhD in Electrical Engineering from the Univeristy of Nottingham in 2013. He is currently a lecturer two (LII) at the FUT Minna, Nigeria, a position which he started in 2005. He is a registered engineer with the Council for the Regulation of Engineering in Nigeria (COREN), a professional member of the Institute of Electrical and Electronics Engineers (IEEE), and a professional member of the Association for Computing Machinery (ACM). His research interests are in the fields of swarm intelligence, routing algorithms, wireless sensor networks, energy harvesting, automation, home and industrial security system.

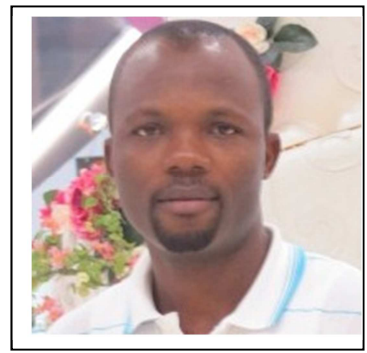

**APPENDIX I**

**Components List**

| RESISTORS | VALUE | RATING |
|---|---|---|
| $R_1$ | 10K | 1/4w |
| $R_2$ | 10K | 1/4w |
| $R_3$ | 10K | 1/4w |
| $R_4$ | 10K | 1/4w |
| $R_5$ | 680K | 1/4w |
| $R_6$ | 470K | 1/4w |
| $R_7$ | 100 | 1/4w |
| $R_8$ | 10M | 1/4w |
| $R_9$ | 10K | 1/4w |
| $R_{10}$ | 4.7K | 1/4w |
| CAPACITORS | VALUE | Rating |
| $C_1$ | 10µF | 16v |
| $C_2$ | 470µF | 16v |
| $C_3$ | 100NF | Standard |
| $C_4$ | 100nF | Standard |
| $C_5$ | 100µF | 25V |

**INTEGRATED CIRCUITS**

I.C1  -   MCI 4043 QUAD 3 – STATE RS LATCH

I.C2  -   +12VD.C. VOLTAGE REGULATOR (7812)

I.C3  -   BRIDGE RECTIFIER (2WO4M)

TRANSISTOR

T1 BC 547

DIODES

D1, D2  IN 4148 SWITCHING DIODES

<u>MISCELLANEOUS</u>

10 PUSH ON/RELEASE OFF MINIATURE BUTTON SWITCHES

1      CENTER TAP MAINS TRANSFORMER WITH 12V SECONDARY WINDING

1      CENTER TAP MAINS TRANSFORMER WITH 15V SECONDARY WINDING